\documentclass{INTERSPEECH2023}
\interspeechcameraready 
\title{Visually-Aware Audio Captioning With Adaptive Audio-Visual Attention}

\name{
      Xubo Liu$^{1,*}$,
      Qiushi Huang$^{1,4,*}$,
      Xinhao Mei$^{1,*}\thanks{* The first three authors contributed equally to this work.}$,
      Haohe Liu$^{1}$,
      Qiuqiang Kong$^{2}$,
      \\Jianyuan Sun$^{1}$,
      Shengchen Li$^{3}$,
      Tom Ko$^{2}$,
      Yu Zhang$^{4}$,
      Lilian H. Tang$^{1}$,
      \\Mark D. Plumbley$^{1}$,
      Volkan Kılıç$^{5}$,
      Wenwu Wang$^{1}$}
      
\address{$^1$University of Surrey, UK  \hspace{0.1cm} $^2$ByteDance, China \hspace{0.1cm} $^3$Xi'an Jiaotong-Liverpool University, China\\
  {$^4$Southern University of Science and Technology, China $^5$Izmir Katip Celebi University, Turkey}
 }
\email{}
\begin{document}

\maketitle
\begin{abstract}

Audio captioning aims to generate text descriptions of audio clips. In the real world, many objects produce similar sounds. How to accurately recognize ambiguous sounds is a major challenge for audio captioning. In this work, inspired by inherent human multimodal perception, we propose visually-aware audio captioning, which makes use of visual information to help the description of ambiguous sounding objects. Specifically, we introduce an off-the-shelf visual encoder to extract video features and incorporate the visual features into an audio captioning system. Furthermore, to better exploit complementary audio-visual contexts, we propose an audio-visual attention mechanism that adaptively integrates audio and visual context and removes the redundant information in the latent space.  Experimental results on AudioCaps, the largest audio captioning dataset, show that our proposed method achieves state-of-the-art results on machine translation metrics.

\end{abstract}

\section{Introduction}
\label{sec:intro}
Automatically generating textual descriptions for audio clips, a task known as automated audio captioning (AAC) \cite{mei2022ac_review}, has attracted increasing attention in recent years. AAC has significant potential in practical applications, such as helping the hearing-impaired to understand sounds, indexing audio in large-scale databases \cite{mei2022metric, lass}, and facilitating conversational AI systems \cite{huang2022personalized}. 

In the real world, many objects produce similar sounds, therefore, discriminating sounding objects with access to audio information only is difficult even for humans. For example, consider the following two human-annotated captions for the same audio clip from the Clotho dataset \cite{drossos2020clotho}: `\textit{There was a jackhammer running at a construction site in the distance}'; and `\textit{A motor vehicle is running with speed and stopped its engine}'. The sound of `\textit{jackhammer}' and `\textit{motor}' are both mechanical sounds, and the human annotators cannot clearly discriminate these ambiguous sounding objects when only audio is available. How to accurately identify these types of ambiguous sounds is a major challenge for audio captioning systems.

Humans perceive the world through hearing and vision, where the auditory and visual cues often provide complementary information \cite{tadas2018mml_review, SynthVSR}. Visual cues can help improve audio perception by recognizing the visual objects that might be potential sounding objects. Visual information would offer great help to the identification of ambiguous sounds for audio captioning systems in many applications where video streams are available, such as generating audio captions for YouTube videos and helping hearing-impaired people access the acoustic environment with multi-modal smart devices. To our knowledge, visual information has not been investigated in the literature to improve the performance of audio captioning.

Motivated by human audio-visual perception, we propose a visually-aware audio captioning method that leverages visual context to improve audio captioning systems. Specifically, an off-the-shelf visual encoder is introduced into an encoder-decoder audio captioning system and processes visual inputs in a similar way to the audio encoder. The incorporation of visual information can also help 
infer sounding objects. For example, consider an audio clip with the sound of a motor, the visually-aware captioning model can easily recognize the appearance of a motor in visual context, while an audio-only captioning model cannot.
However, video frames are highly redundant \cite{nagrani2021attention} for auditory perception as generally not too many visual objects make sounds in a video clip. The direct incorporation of visual information introduces redundant information, thereby degrading the audio captioning system, as shown in our experimental results in Section 4.5. To address this issue, we further propose an adaptive audio-visual attention mechanism to exploit complementary contexts in audio and visual modalities and simultaneously
drop redundant audio-visual information in the latent feature space. We conduct experiments on AudioCaps \cite{kim2019audiocaps}, the largest public audio captioning dataset. The experimental results show that the proposed approach significantly improves the performance of an off-the-shelf audio captioning baseline system \cite{Mei2021ACT}, indicating the effectiveness of introducing visual context to audio captioning, and our proposed system achieves the state-of-the-art results on machine translation metrics. Our code is available at \url{https://github.com/liuxubo717/V-ACT}.

The remainder of this paper is organized as follows. Section 2 introduces the related work. Section~\ref{sec:method} introduces the method we proposed for visually-aware audio captioning. Section~\ref{sec:exp} presents the experiments and results. We conclude this work and discuss the future direction in Section~\ref{sec:conclusion}.

\begin{figure*}[!t]
  \centering
  \vspace{-2em}
  \includegraphics{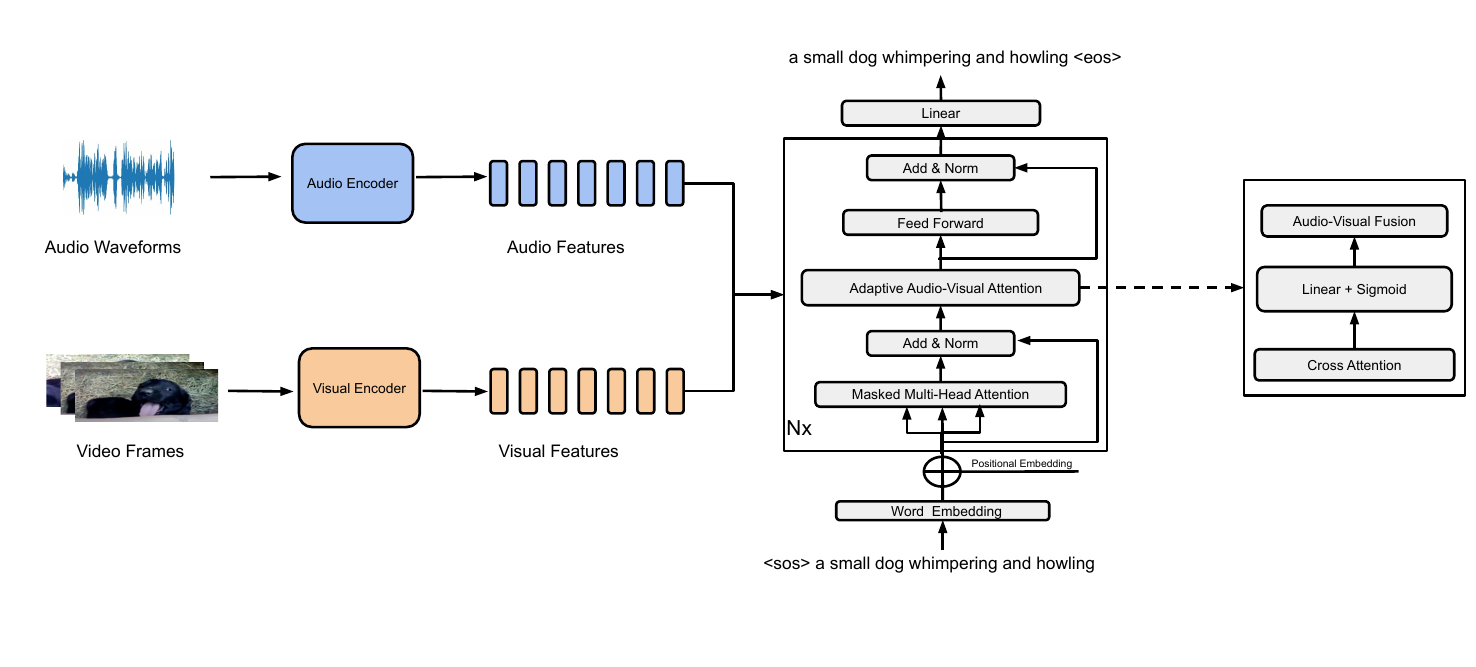}
  \vspace{-3em}
  \caption{Overview of the proposed visually-aware audio captioning system. A visual encoder is introduced to encode visual features and an adaptive audio-visual attention block is used to replace the cross-attention block in the text decoder of the conventional audio-only captioning system, in order to fuse the visual information with audio.}
  \label{fig:system_overview}
\vspace{-1em}
\end{figure*}

\section{Related work}
\label{sec:related_works}

\subsection{Audio captioning}
Audio captioning is generally framed as a cross-modal translation task \cite{mei2022ac_review}, and the encoder-decoder framework is widely adopted, where the encoder extracts latent audio feature representations from input audio, which are then fed to the decoder for generating captions. Recently, many efforts have been made to investigate the effect of different network architectures as the encoders and decoders, among which Convolutional Neural Networks (CNNs) and Transformers show superior performance to others as the audio encoders while RNNs and Transformers are the most popular text decoders \cite{Mei2021ACT, chen2020ac_CNN, Mei2021ac_trans}. In addition to investigations into network architectures, auxiliary information such as fusion of audio features \cite{sun2022automated}, keywords or sentence information \cite{koizumi2020keywords}, different training strategies such as contrastive learning \cite{chen2022contrastive, Liu2021cl4ac}, reinforcement learning \cite{Mei2021ac_trans} and generative adversarial training \cite{mei2021diverse}, have been applied to improve the performance of AAC systems. Existing works in AAC only consider audio information. In this paper, we show a first attempt to leverage visual context to improve  the performance of AAC systems when video streams are available. 

\subsection{Audio-visual learning}
Recently, there has been a surge of research interest in audio-visual tasks. For example, audio information has been exploited in video captioning, video retrieval, and visual question answering \cite{tian2018av_v_vap, Seo2022pre_mm_v_cap}, with audio-visual multi-modal configurations generally outperforming uni-modal video-only configurations \cite{Iashin2020mm_v_cap}. Different approaches were investigated on how to effectively learn, align and fuse audio and video modalities \cite{nagrani2021attention}. Most of these works were applied in the video domain, however, little attention has been paid to using video information to improve audio-oriented tasks, such as audio event classification \cite{shirian2022visually, boesh2021av_transfer}.

Our task is audio captioning which is close to audio-visual video captioning \cite{Seo2022pre_mm_v_cap, Iashin2020mm_v_cap}, especially after we introduce visual information. However, audio-visual video captioning focuses on recognizing and describing visual objects and their actions, with reference to the audio streams, while audio captioning aims to describe only the audio context. In many audio captioning applications such as generating audio captions for hearing-impaired or YouTube videos, users do not need the visual context in the captions as they can directly access the visual world, and using audio-visual captions will introduce redundant information especially when there are many unsounding visual contexts. In this work, we focus on exploring the potential of introducing visual information to help identify ambiguous sounds and thus improve the performance of audio captioning systems.   

\section{Proposed Method}
\label{sec:method}
Our proposed visually-aware audio captioning method, as shown in Figure~\ref{fig:system_overview}. This method is built on the audio captioning transformer (ACT) baseline developed in \cite{Mei2021ACT}, by incorporating a visual encoder along with the audio encoder, and an adaptive audio-visual attention mechanism into the text decoder. 

\subsection{Auditory and Visual Encoding}
Audio and visual encoders take audio waveforms and video frames as input, and generate audio features and visual features as output, respectively. The ACT audio encoder is a Transformer-based neural network adapted from Vision Transformer (ViT) \cite{dosovitshiy2021vit} and consists of \num{12} identical blocks, each of which contains a self-attention block and a multi-layer perceptron block (MLP). A mel-spectrogram is first extracted from the audio waveforms and then split into non-overlapping patches along the time axis. These patches are then projected into a latent space via a linear layer and added with a learnable positional embedding before they are fed into the main encoder block. To address the scarcity issue of paired audio-text data, the ACT encoder is first pre-trained on AudioSet \cite{audioset}, the largest audio event dataset, with an audio tagging task. 

For the visual encoder, we adopt the video CNN-separable conv3D (S3D) \cite{xie2017s3d} pre-trained on kinetics, a large-scale human action video dataset \cite{kay2017kinetics}. S3D is an improvement over I3D \cite{carreira2017i3d}, which is a 3D Inception architecture. It replaces the standard 3D convolutions with spatiotemporal-separable 3D convolutions, resulting in higher accuracy and computational efficiency when working on video-related tasks. A linear layer is appended after both encoders to map the audio and visual features into the same dimension. It should be noted that our proposed method is agnostic to the network architecture of the audio and visual encoders, and any suitable neural networks could be employed for encoders. We freeze the video encoder during training to save memory and running time. 

\subsection{Adaptive Audio-Visual Attention}
In ACT \cite{Mei2021ACT}, the text decoder is a standard Transformer decoder \cite{vaswani2017attention} consisting of a masked self-attention block, a cross-attention block, and an MLP block. 
The text decoder generates captions word by word in an auto-regressive manner. Suppose we are going to generate the $t$-th word in a caption and the previously generated $t-1$ words are denoted as $w_{1:t-1}$. The previously generated words are first passed through a word embedding layer to get the word embeddings $W \in \mathbb R^{(t-1) \times d_w}$, where $d_w$ is the dimension of the word embeddings. The word vectors are then sent to the masked self-attention layer with a residual connection, this can be formalized as:
\begin{equation}
\begin{aligned}
H_{\rm{attn}} &= \rm{SelfAttention} (\textit{W}) \\
H_{\rm{hidden}} &= H_{\rm{attn}} + W \\
\end{aligned}
\end{equation}
where $H_{\rm{hidden}} \in \mathbb R^{t-1 \times d}$ is the output of the self-attention block. 

Visual information is often redundant across frames, with some visual objects lacking accompanying sound. To address this challenge, we design adaptive audio-visual attention, which can effectively exploit contextual cues from both modalities.
Suppose an audio feature sequence $A \in \mathbb R^{T_{\rm{a}} \times d}$ and a video feature sequence $V \in \mathbb R^{T_{\rm{v}} \times d}$ are extracted by the audio encoder and visual encoder respectively, where $T_a$ is the number of audio features, $T_v$ is the number of visual features, and $d$ is the dimension of the features. The feature sequences $A$ and $V$, together with the self-attention output $H_{\rm{hidden}}$, are first sent to a cross-attention layer to calculate the audio and visual cross-attention outputs $A_{\rm{cross}}$ and $V_{\rm{cross}}$, through which text information is fused with audio and visual information, respectively. This can be formalized as
\begin{equation}
\begin{aligned}
    A_{\rm{cross}}= {\rm{CrossAttention}} (H_{\rm{hidden}},A) \\
    V_{\rm{cross}}= {\rm{CrossAttention}} (H_{\rm{hidden}},V) \\
\end{aligned}
\end{equation}
where $A_{\rm{cross}}$ and $V_{\rm{cross}}$ have the same shape as $H_{\rm{hidden}}$. In this layer, $A$ and $V$ act as both keys and values, while $H_{\rm{hidden}}$ acts as queries in cross-attention computation. Afterward, $A_{\rm{cross}}$ and $H_{\rm{hidden}}$ will be concatenated and fed into a linear layer with a sigmoid activation to get the confidence score $A_{\rm{conf}} \in \mathbb R^{t-1 \times d}$ with respect to the audio information. Formally:

\begin{equation}
    A_{\rm{conf}} = {\rm{Sigmoid}} ({\rm{FC}} ([A_{\rm{cross}};H_{\rm{hidden}}]))
\end{equation}
where $\rm{FC}$ is a fully-connection layer, $\rm{Sigmoid}$ is a sigmoid activation function and $[;]$ represents the concatenation operation. After that, a hyperparameter confidence threshold $\beta$ is introduced to generate masks $M_a$ for audio information $A_{\rm{cross}}$ and $M_{\rm{v}}$ for visual information $V_{\rm{cross}}$ based on the confidence score $A_{\rm{conf}}$.   
\begin{equation}
\begin{aligned}
    M_{\rm{a}} &= {\rm{Mask}}(A_{\rm{conf}},\beta)\\
    M_{\rm{v}} &= {\rm{Mask}}(1 - A_{\rm{conf}}, \beta)\\
\end{aligned}
\end{equation}
where the mask function is defined as:
\begin{equation}
{\rm{Mask}}(X, \beta)=
\begin{cases}
1,  &\rm{if}\ X_{i, j} > \beta \\
0,  &\rm{otherwise}. 
\end{cases}
\end{equation}
Finally, after getting the masks, $A_{\rm{cross}}$ and $V_{\rm{cross}}$ are fused to get the output $AV_{\rm{out}} \in \mathbb R^{t-1 \times d}$ in the final layer of this adaptive audio-visual attention block:
\begin{equation}
    AV_{\rm{out}} = A_{\rm{conf}} \odot A_{\rm{cross}} \odot M_{\rm{a}} + (1-A_{\rm{conf}}) \odot V_{\rm{cross}} \odot M_{\rm{v}}.
\end{equation}
where $\odot$ is the Hadamard product. No residual connection is employed in this block, thus $AV_{out}$ will be directly fed into the next feed-forward layer in the decoder block. 

In summary, the proposed audio-visual attention mechanism can adaptively exploit complementary context from visual and acoustic streams while eliminating the audio-visual redundancy in the latent feature space. In addition, our proposed method also serves as a regularization to the model that can avoid overfitting as we produce sparse audio-visual features via a masking strategy. We demonstrate the superiority of our proposed method in the part of the experiment.

\section{Experiments}
\label{sec:exp}

\begin{table*}[ht]
\caption{Scores on the AudioCaps test set. ACT: uni-modal audio captioning baseline. V-ACT: Visually-aware Audio Captioning Transformer that introduces an off-the-shelf visual encoder to leverage visual information. AdaAVA: proposed adaptive audio-visual attention. AdaAVA-audio: confidence score is calculated based on the audio cross attention $A_{\rm{cross}}$. AdaAVA-video: confidence score is calculated based on the video cross attention $V_{\rm{cross}}$. V-ACT (video-only): uni-modal video captioning baseline. V-ACT (concatenate): baseline system using concatenate-based audio-visual fusion. The highest value for each metric is shown in bold.}
\centering
\begin{tabular}[\linewidth]{c c c c c c c c c c c} 
 \hline
 Model & BLEU$_{1}$ & BLEU$_{2}$ & BLEU$_{3}$ & BLEU$_{4}$ & ROUGE$_{L}$ & METEOR & CIDEr & SPICE & SPIDEr \\ 
 \hline
 RNN+RNN \cite{kim2019audiocaps} & 0.614 & 0.446 & 0.317 & 0.219 & 0.450 & 0.203 & 0.593 & 0.144 & 0.369 \\
 CNN+RNN \cite{xu2021invest_cnn_crnn} & 0.655 & 0.476 & 0.335 & 0.231 & 0.467 & 0.229 & 0.660 & 0.168 & 0.414 \\
 CNN + BERT \cite{liu2022leveraging} & 0.671 & 0.498 & 0.358 & 0.251 & 0.480 & 0.232 & 0.667 & 0.172 & 0.419 \\
 CNN+Transformer \cite{Mei2021ac_trans} & 0.641 & 0.479 & 0.344 & 0.236 & 0.469 & 0.221 & 0.693 & 0.159 & 0.426 \\
 BART + AudioSet tags \cite{gontierautomated} & \textbf{0.699} & 0.523 & 0.380 & 0.266 & 0.493 & \textbf{0.241} & \textbf{0.753} & \textbf{0.176} & \textbf{0.465} \\
 \hline
 ACT (audio-only) \cite{Mei2021ACT} & 0.685 & 0.518 & 0.376 & 0.263 & 0.488 & 0.233 & 0.678 & 0.169 & 0.424 \\
V-ACT (video-only)  & 0.548 & 0.365 & 0.238 & 0.153 & 0.381 & 0.168 & 0.403 & 0.114 & 0.259 \\
V-ACT (concatenate)  & 0.687 & 0.509 & 0.359 & 0.239 & 0.490 & 0.228 & 0.675 & 0.161 & 0.418 \\
 V-ACT (AdaAVA-video) & 0.691 & 0.515 & 0.373 & 0.261 & 0.489 & 0.230 & 0.682 & 0.168 & 0.425  \\
 V-ACT (AdaAVA-audio) & 0.698 & \textbf{0.527} & \textbf{0.388} & \textbf{0.281} & \textbf{0.494} & 0.237 & 0.711 & 0.172 & 0.442  \\
\hline
\end{tabular}

\label{tab:results} 
\end{table*}

\subsection{Dataset}
We conduct experiments on AudioCaps \cite{kim2019audiocaps} whose audio clips are sourced from YouTube, so its corresponding video clips can be also downloaded for research purpose\footnote{Other audio captioning datasets, such as Clotho, do not have video sources.}. AudioCaps \cite{kim2019audiocaps} is the largest public audio captioning dataset with around 50k 10-second audio clips, and is divided into three splits: training, validation, and testing sets. The audio clips are annotated by humans through the Amazon Mechanical Turk (AMT) crowd-sourced platform. Each audio clip in the training sets has a single human-annotated caption, while each clip in the validation and test set has five human-annotated captions. Because some video clips are no longer available on YouTube, our downloaded version contains \num{47745} audio clips in the training set, \num{480} clips in the validation set, and \num{928} clips in the test set.

\subsection{Implementation details}
For data pre-processing, all the captions are converted to lowercase with punctuation removed, and two special tokens are appended at the start and end of each caption. The tokens are split based on the word level. Audio clips are sampled with \num{32}kHz. A \num{1024}-point Hanning window with \num{640}-point hop size and \num{64} mel bins are used to extract mel-spectrograms, which results in mel-spectrograms with a shape of $1000 \times 64$. The split patch size is $4 \times 64$. Each video clip is uniformly segmented into non-overlapping \num{250}-frame chunks for video feature extraction. The dimension of the audio and visual features is set to \num{512}.

ACT-m~\cite{Mei2021ACT} is used as the baseline whose decoder contains \num{4} Transformer blocks and \num{8} heads. The model is trained with Adam optimizer for \num{15} epochs with a learning rate of \num{1e-4}, and linear warmup is applied in the first \num{5} epochs. Batch size is set to \num{32}. The confidence threshold $\beta$ is set to \num{0.13} by hyperparameter searching. Label smoothing and Specaugment~\cite{park2019specaugment} are also employed to avoid overfitting. During inference, a beam search with a beam size up to \num{3} is used for decoding.

\subsection{Evaluation metrics}
We evaluate our proposed model using conventional automatic metrics \cite{mei2022ac_review} borrowed from machine translation and image captioning, mostly are calculated based on $n$-gram matching ($n$-gram refers to $n$ consecutive words). BLEU$_n$ measures the precision of $n$-gram matching and a sentence-brevity penalty is introduced to penalize short sentences. ROUGE$_l$ calculates an F-measure by considering longest common subsequence between the candidate and ground-truths. METEOR calculates uni-gram precision and recall, taking into account the surface forms, stemmed forms, and meanings of words. CIDEr computes the cosine similarity of weighted $n$-grams between candidates and references. SPICE parses each caption into scene graphs and an F-measure is calculated based on the matching of the graphs. SPIDEr is the average of SPICE and CIDEr. BLEU$_n$, ROUGE$_l$, and METEOR are machine translation metrics and CIDEr, SPICE, SPIDEr are image captioning metrics.
\subsection{Results}
Table~\ref{tab:results} presents our experimental results. We first compare the results of our proposed method with those of different uni-modal baselines. For the uni-modal (audio-only) architectures, we observe that our ACT baseline shows similar performance with the CNN+RNN \cite{xu2021invest_cnn_crnn}, CNN+BERT \cite{liu2022leveraging} and CNN+Transformer \cite{Mei2021ac_trans} models, while significantly outperforms the RNN-RNN model \cite{kim2019audiocaps}. The result of our proposed visually-aware audio captioning model with adaptive audio-visual attention i.e. \textit{V-ACT (AdaAVA-audio)} is shown in the last row of Table~\ref{tab:results}. The scores on all evaluation metrics are significantly improved and higher than all the uni-modal configurations. This suggests that leveraging visual information for describing audio events is effective. Furthermore, we compare the performance of our method with the state-of-the-art method \cite{gontierautomated} that uses additional AudioSet tags information. Our method is on par with the state-of-the-art, specifically, we achieve the state-of-the-art on machine translation metrics BLEU$_{2-4}$ and ROUGE$_l$, indicating that our proposed system ensures better syntactic fluency and semantic similarity. It should be noticed that the state-of-the-art method \cite{gontierautomated} is highly dependent on leveraging the additional text information i.e. AudioSet tags which are used as word hints in the caption annotation stage of AudioCaps dataset \cite{kim2019audiocaps} creation procedure. This method may not generalize well to other audio captioning datasets \cite{liu2022leveraging}.

\subsection{Ablation studies}
We further conduct experiments to study the contribution of the visual information and the effect of the proposed audio-visual attention mechanism. Firstly, when only using the video inputs, that is, the audio encoder is not used and the system just contains a visual encoder and a caption decoder, which is defined as \textit{V-ACT (video-only)} in Table 1. Video-only method performs the worst in all metrics as compared to other models. The results we obtained are consistent with the previous work \cite{shirian2022visually} on visually-aware sound event classification. As the context in video modality may be irrelevant to sound events~(i.e., visual objects do not make sound). In addition, sounding objects may be out of the screen. Therefore, only using video inputs is insufficient for captioning. This also indicates the gap between audio captioning and video captioning, where the former concentrates on audio events while the latter focuses more on the actions of the visual objects. Secondly, instead of fusing the audio and visual information through our proposed attention mechanism, we also evaluated the system by directly concatenating the audio and visual features $A$ and $V$ obtained from the audio encoder and video encoder, respective, as defined as \textit{V-ACT (concatenate)} shown in third last row in Table~\ref{tab:results}. It can be observed that directly concatenating the audio and visual features does not improve the system performance, and the metrics are on par with or even slightly lower than the uni-modal ACT baseline. This supports our previous claim that visual information is generally redundant and simply fusing audio and visual information may degrade the audio captioning performance. On the other hand, our proposed adaptive audio-visual attention mechanism could adaptively fuse audio and visual features based on the confidence of audio information, thereby avoiding degraded fused representation caused by visual interference. Finally, we investigate the performance of the system \textit{V-ACT (AdaAVA-video)} whose confidence score $A_{\rm{conf}}$ is calculated based on the video cross attention, formally: 
\begin{equation}
    A_{\rm{conf}} = {\rm{Sigmoid}} ({\rm{FC}} ([V_{\rm{cross}};H_{\rm{hidden}}]))
\end{equation}
The results are shown in the penultimate row in Table 1. \textit{V-ACT (AdaAVA-video)} performs worse than the \textit{V-ACT (AdaAVA-audio)}, we analyze this because 1) audio captioning task itself is an audio-oriented task, the video stream is not representative for audio context 2) video feature is pre-processed which restricts the learning ability of the system.

\section{Conclusion}
\label{sec:conclusion}
In this work, we proposed visually-aware audio captioning, which aims to leverage visual information to help recognize ambiguous sound and improve caption accuracy. As the visual information for in-the-wild video is often noisy and redundant, an adaptive audio-visual attention mechanism is designed to adaptively exploit complementary audio-visual context and eliminate redundant information. Experimental results show that the proposed method outperforms audio-only uni-modal models and achieves state-of-the-art on machine translation metrics. Further research should be carried out to improve the efficiency and robustness of the proposed model, especially after introducing a state-of-the-art video encoder \cite{neimark2021video}.

\section{Acknowledgement}
\label{sec:ack}
This work is partly supported by UK Engineering and Physical Sciences Research Council (EPSRC) Grant EP/T019751/1 ``AI for Sound'', a Newton Institutional Links Award from the British Council, titled ``Automated Captioning of Image and Audio for Visually and Hearing Impaired" (Grant number 623805725), British Broadcasting Corporation Research and Development~(BBC R\&D), a PhD scholarship from the University of Surrey, and a Research Scholarship from the China Scholarship Council (CSC). For the purpose of open access, the authors have applied a Creative Commons Attribution (CC BY) licence to any Author Accepted Manuscript version arising.
\newpage

\bibliographystyle{IEEEtran}
\bibliography{refs, strings}

\end{document}